\documentclass{emulateapj}

\newcommand{\deltE}{\Delta\kern-1ptE}

\input epsf

\usepackage{longtable}
\usepackage{amssymb}
\usepackage{graphicx}
\usepackage{epsfig}
\usepackage{epsf}

\slugcomment{Accepted by ApJ}
\shorttitle{X-ray irradiated protoplanetary disk atmospheres}
\shortauthors{Ercolano et al.}
\begin{document}
\title{X-ray irradiated protoplanetary disk atmospheres I: 
Predicted emission line spectrum and photoevaporation}
\author{Barbara Ercolano$^{1,2}$, Jeremy J. Drake$^2$, John C. Raymond$^2$, Cathie C. Clarke$^1$}
\affil{$^1$Institute of Astronomy, University of Cambridge, \\Madingley Rd, \\Cambridge, CB3 OHA, UK\\
$^2$Harvard-Smithsonian Center for Astrophysics, MS-67
\\ 60 Garden Street, \\ Cambridge, MA 02138, USA}

\begin{abstract}

We present {\sc mocassin} 2D photoionisation and dust radiative
transfer models of a prototypical T~Tauri disk irradiated by X-rays
from the young pre-main sequence star. The calculations demonstrate a
layer of hot gas reaching temperatures of $\sim$10$^6$K at small radii
and $\sim$10$^4$K at a distance of 1~AU. The gas temperatures decrease
sharply with depth, but appear to be completely decoupled from dust
temperatures down to a column depth of
$\sim5\times$10$^{21}\,cm^{-2}$.

We predict that several fine-structure and forbidden lines of heavy
elements, as well as recombination lines of hydrogen and helium,
should be observable with current and future
instrumentation, although optical lines may be smothered by the stellar 
spectrum. Predicted line luminosities are given for the the
brightest collisionally excited lines (down to
$\sim$10$^{-8}$L$_{\odot}$), and for recombination transitions from
several levels of H~{\sc i} and He~{\sc i}.

The mass loss rate due to X-ray photoevaporation estimated from our
models is of the order of $10^{-8}$~M$_{\odot}$ yr$^{-1}$, implying a
dispersal timescale of a few Myr for a disk of mass 0.027~M$_{\odot}$, which is 
the mass of the disk structure model we employed.
We discuss the limitations of our model and highlight the need for
further calculations that should include the simultaneous solution of
the 2D radiative transfer problem and the 1D hydrostatic equilibrium
in the polar direction.

\end{abstract}


\section{Introduction}\label{s:intro}

Protoplanetary disks are a product of the formation of protostars
from the collapse of molecular cloud cores.  These disks remain a
source of ongoing accretion for the central stellar object for a
period of up to $\sim 10$~Myr (Manning \& Sargent, 1997; Wyatt, 
Dent \& Greaves, 2003) and are the birthplaces of planetary
systems.  The evolution of the disk and of its planetary progeny is
controlled by heating and irradiation from the central
star---from IR to X-rays---but in the vicinity of OB stars 
can also be significantly affected by the external radiation environment 
(e.g. Johnstone et al. 1998; Adams et al., 2004; Fatuzzo \& Adams, 2008) .   

The discovery of the ``proplyd'' phenomenon and subsequent
observations of protoplanetary disks represent an important and
growing part of the legacies of the {\it Hubble} and {\it Spitzer}
space telescopes (e.g. O'Dell 1993, Bally et al. 1998, Evans et al.
2003, Meyer et al. 2004).  New instruments such as the {\it
Mid-Infrared Instrument} on board {\it James Webb Space Telescope},
which will provide both imaging and spectroscopy, hold the promise for
higher spatial resolution.  On the ground, millimeter and
submillimeter studies have evolved from examining disk size and
orientation (e.g. Keene \& Masson 1990, Lay et al. 1994) to detailed
studies of structure, chemistry and thermal properties (e.g. Wilner
et al. 2003, Qi et al.\ 2006), while future high spatial resolution
submillimeter observations with ALMA promise further major advances
(e.g. van Dishoeck \& J\"orgensen 2008). 

These and other observations should provide insights into issues
central to understanding disks and the likelihood of forming planetary
systems like our own, such as angular momentum transport, and the
timescales for grain growth and gas dissipation.  Of particular
interest for interaction of the disk with its parent star and for the
study of planetary origins are the inner regions of disks ($\le
10$~AU) where both terrestrial and giant planets are believed to form.

While the observational database continues to grow, all attempts at
theoretically modeling the emission expected from such environments
have lacked several of the ingredients needed for a fully
self-consistent approach. The multidimensional, far-Ultra Violet (UV)
to X-ray radiative transfer (RT) problem must be solved simultaneously
with the disk hydrodynamical evolution, to which it is strongly
coupled. Furthermore, disks comprise adjacent regions of ionised,
photon-dominated and molecular gas intermixed with a non-uniform dust
component.  No computer code currently available includes all the
important physics and chemistry needed to properly interpret the
observational diagnostics of these different regimes.  Furthermore,
full theoretical understanding of a number of important contributors
to the energetics of the system, such as accretion and wind-disk
interactions, is still lacking; these ingredients can currently only be
included through approximate and phenomenological approaches (e.g.\
see discussion in Glassgold et al., 2004).

Several studies have focused on the properties of the dust component
of disk atmospheres (e.g. D'Alessio et al. 2001, Dullemond et
al. 2002, Akeson et al 2005). However, studies of the gaseous
component of the inner disks have only recently been attempted, and
only an exploratory picture of the thermochemical structure of this
crucial component has emerged.  Pioneering work from Igea \& Glassgold
(1999), Glassgold et al., (2004, 2005) and Meijerink et al. (2008)
examined physical conditions and expected emission lines in the
gaseous component of non-evolving T~Tauri disks irradiated by X-rays.
Nomura \& Millar (2005) modeled molecular hydrogen emission from
protoplanetary disks, taking into account UV radiation from a central
star; their models, while calculating the density and temperature
structures self-consistently, did not account for X-radiation
and employed an approximate treatment of the scattered light that
often leads to an overestimation of the UV radiation field in the disk
(Glassgold, 2006).  Semenov et al. (2004) employed perhaps
the most complete chemical network to date to calculate the ionisation
fractions at the disk midplane.  While X-ray ionisation was included,
radiative transfer through scattering and diffusion was not.  Simple
column density arguments were used instead to estimate the radiation
field attenuation at discrete points in the disk where the chemical
model was applied. Very recently Gorti \& Hollenbach presented 1+1D models 
of gas in optically thick disks, calculating thermal, density and 
chemistry structures, with special attention to the effects of FUV irradiation. 

More detailed 2D or 3D calculations are now needed to provide
a quantitative picture of these environments in relation to their
effects on the disk spectral energy distribution (SED),
and on other observables available now and in
the near future (e.g. $[$O~{\sc i}$]$, $[$Ne~{\sc ii}$]$, CO, OH, and
UV transitions of H$_2$; Najita et al, 2007).  
In particular, observations of forbidden lines in T~Tauri stars
provide evidence for a tenuous, hot outer layer, or ``corona' (Kwan \&
Tademaru 1988, 1995; Kwan (1997).  The earlier work by by Glassgold et
al. (2004, 2007), Semenov et al. (2004) and Meijerink et al. (2008)
emphasised the chemistry in the photo-dissociation region (PDR) and
molecular zones, and stopped short of this coronal layer. Also the 
recent work by Gorti \& Hollenbach does not include the fully ionized 
layer discussed here.

In this paper, which is the first of a series aimed at building a more
realistic and self-consistent model of irradiated gaseous T~Tauri
disks, we employ a fully 3D photoionisation and dust radiative
transfer code, {\sc mocassin} (Ercolano et al., 2003, 2005, 2008), to
calculate the detailed ionisation and temperature structure of a
typical T~Tauri disk, with special emphasis on the
photoionisation-dominated outer layers and corona.  We show that hot
coronal temperatures are expected from X-ray irradiation from the
central pre-main sequence star. Our calculations aim to understand the
thermochemical structure of the irradiated region and to identify
emission lines that can be used as gas-phase diagnostics for current
and future observations.  Furthermore, the detailed 2D temperature
structure calculated by our models allows us to estimate 
the efficiency of X-radiation to drive a photoevaporative wind.

Our model is described in Section 2. The ionisation and temperature
structures are presented and discussed in Section 3, while the
predicted emission line spectra are shown in Section 4. Section 5
contains a discussion of our results in the context of disk dispersal
via X-ray photoevaporation. A brief summary is given in Section 6.

\section{The 2D Monte Carlo photoionisation and dust radiative
  transfer model}\label{s:model}

Photoionisation and temperature structure calculations were
performed with Version 3.02.00 of the 3D Monte Carlo photoionisation
and dust radiative transfer code, {\sc mocassin} (Ercolano et
al. 2003, 2005, 2008). The code uses a stochastic approach to the
transfer of radiation allowing for the self-consistent transfer of
both the primary and secondary components of the radiation field.
The code was modified to include viscous heating, approximated using
the prescription for a thin disk by Pringle (1981), and gas cooling by
collisions of grains with a mixture of atomic and molecular hydrogen
(Hollenbach \& McKee, 1979); we discuss this further below. 
{\sc mocassin} was also adapted
to run in 2D, using the gas and dust
density distribution given by a 2D
protoplanetary disk calculation of D'Alessio et al. (1999).


The D'Alessio model used in our calculation was chosen to be that
which best fits the median SED of T~Tauri stars in Taurus (d'Alessio, 2005).
 We refer the reader to D'Alessio et al.\ (1999) for a
detailed description of the model ingredients and calculation.  In
brief, the input parameters for this model include a central star of
0.7~M$_{\odot}$, 2.5~R$_{\odot}$ irradiating the disk with an
effective temperature of 4000~K. Additional disk parameters consist of
a mass accretion rate of 1.e-8 M$_{\odot}$ yr$^{-1}$ and a viscosity
parameter $\alpha$~=~0.01. The total mass of the disk is  0.027~M$_{\odot}$.

The present study is particularly focused on determining the physical
properties of the gas in the outer layers and hot disk
corona. Molecular species that are expected to form deeper in the disk
are not included in our model; this poses a limitation on the accuracy
of our gas temperature calculations in the inner layers. In
particular, the calculations by Glassgold et al., (2004, GNI04) have
shown that CO rovibrational lines may become important coolants, and,
e.g., at a radial distance of 1~AU and column densities of
10$^{21}$-10$^{22}$~cm$^{-2}$ CO was found to dominate the disk cooling.  CO
rotational lines are also expected to form at greater depths, though
in these regions the thermal balance is dominated by dust-gas
collisions (GNI04).
 
Dust-gas collisions can heat the gas/dust if the gas is cooler/warmer
than the dust; photoelectric emission from dust grains heats the gas
and it provides a minor cooling channel for the dust. {\sc mocassin}
is able to self-consistently calculate the dust and gas thermal
balance, taking into account the main microphysical processes that
couple the two phases. Here, however, we choose to keep the dust
temperatures fixed to the values calculated by D'Alessio et al.\
(1999) and focus our discussion on the gas phase. In the current work
we do not treat the transfer of the stellar or interstellar far-UV
field, and therefore the dust temperatures we would obtain from our
models would be incorrect.  The effects of gas-dust interactions
on the temperature
structure of the gas, however, are still taken into account in our gas
thermal balance as well as in the radiative transfer, where the
competition between dust and gas for the absorption of X-ray radiation
is properly treated.

We illuminate the disk using synthetic X-ray spectra representative of
the high energy emission from a typical T~Tauri star. There are
currently at least three different mechanisms by which T~Tauri stars are 
thought to produce observable X-rays: magnetospheric accretion shocks,
shocks resulting from polar jets, and magnetically-confined hot
coronae analogous to those found on all late-type main-sequence stars
including the Sun (e.g. Kastner et al 2002, G\"udel et al 2005, Walter
\& Kuhi, 1981).  The first two of these have only been identified in a
small handful of stars, and since the latter appears to dominate the
observed X-ray spectra of T~Tauri stars we currently ignore the
effects of X-rays from jets and accretion.

Coronal X-ray emission arises from an optically-thin plasma dominated
by impact excitation and ionization by thermal electrons.  The spectra
from such a plasma comprises significant contributions from
bound-bound, bound-free and free-free radiation.  We computed
synthetic coronal spectra incorporating these processes for the energy range
13.6~eV-12.5~keV for all
elements with atomic number Z=1 to 30 using line and continuum
emissivities from the CHIANTI compilation of atomic data (Landi et
al., 2006, and references therein), together with ion populations from
Mazzotta et al. (1998), as implemented in the PINTofALE IDL software
suit\footnote{Freely available from http://hea-www.harvard.edu/PINTofALE} 
(Kashyap \& Drake, 2000). We adopted the solar chemical composition of Grevesse \& Sauval
(1998) for the calculated spectra and an X-ray temperature,
$\log(T_x)=7.2$. We assume an X-ray luminosity,
$L_X$(0.1-10~keV)$=2\times10^{30}$~erg/sec.  

Details on atomic data and
physical processes used for the photoionisation calculations are given
by Ercolano et al. (2003, 2008) and Ercolano \& Storey (2006). The chemical composition of Grevesse \&
Sauval (1998) was also used for the gas abundances of the most common
elements used in our photoionisation model of the disk
corona. In view of the ongoing debate on the ``solar oxygen crisis''
(e.g. Ayres et al. 2006; Drake \& Testa, 2005; Socas-Navarro \&
Norton, 2007; Socas-Navarro \& Centeno, 2008), we have chosen not to
use the more recent compilations of solar abundances based on 3-D
non-LTE hydrodynamic photospheric modeling (e.g. Asplund et al. 2005),
which recommends 25-35\% lower C, N, O and Ne abundances compared to
earlier assessment.  We also ignore the possible effects of grain depletion.
Although some depletion of refractory elements onto dust grains
is expected, the amounts are very uncertain and probably quite
small in the disk corona where the X-ray field is more intense. Here,
undepleted values of C, Mg, Si and Fe abundances correspond to
efficient destruction of grains in the upper layers of the disk. A
test for the validity of this assumption could be provided by the
Si~{\sc ii}~34.8$\mu$m line, which is a good diagnostic for grain
depletion, and which our models predict to be observable (see
Table~1).  

The dust absorption and scattering coefficients are calculated for 
high energies using the dielectric constants for graphite and silicates
of Laor \& Draine (1993), which extend to the X-ray domain.  We assume
spherical grains and use standard Mie scattering series expansion for
$x|m| < 1000$, where $m$ is the complex refractive index and
$x~=~2~\cdot~a/\lambda$ is the scattering parameter (see Laor \&
Draine, 1993). For $x|m| > 1000$ and $x|m-1| < 0.001$ we use
Rayleigh-Gans theory (Bohren \& Huffman, 1983), and for $x|m| > 1000$
and $x|m-1| > 0.001$ we use the treatment specified by Laor \& Draine
(1993), which is based on geometric optics.  

Our dust
model consists of a typical ISM mixture of graphite and silicates with 
MRN size distribution (Mathis, Rumpl \& Nordsiek, 1977) described by 
$a_{min}= 0.005\mu$m and $a_{max}= 0.25\mu$m. The chosen grain
size distribution does not take into account grain growth, and
therefore over-emphasizes the cooling of the gas by the dust in the
disk interior. We also assume a dust to gas mass ratio of 0.01
throughout the disk. We note that the density distribution model of
D'Alessio et al. (1999) uses different dust prescriptions at different
heights, in order to account for the effects of grain settling and
growth. For the sake of simplicity, and because we
assume the dust temperatures calculated by D'Alessio et al. (1999), we
chose to use a more standard dust model throughout our disk. 
Since matching a particular SED is not one of our main objectives, and we
are mostly interested in the physical properties of the disk corona,
the choice of the size distribution and species is not crucial in our
work as it will only have a very small effect on the ionisation and
temperature structure of the coronal gas.

As noted earlier, in addition to all the standard heating and cooling
channels included in {\sc mocassin}, we also include viscous accretion
heating calculated assuming the standard
dissipation rate for a thin disk (Pringle, 1981)
\begin{equation}
\Gamma_{acc} = \frac{9}{4}\alpha\rho c_s^2 \Omega
\end{equation}
where $\rho$ is the local mass density of the gas, $c_s$ is the
isothermal sound speed and $\Omega$ is the angular rotation speed. The
parameter $\alpha$~=0.01 relates viscous heating to the gas pressure.

The formulation above is purely phenomenological and carries large
uncertainties.  In particular, the appropriate value of $\alpha$ is
unclear when the formula is applied to the warm disk atmosphere (GNI04), where 
it is unknown if angular momentum transport such as the magneto-rotational 
instability (MRI; Balbus \& Hawley, 1991) even operates at all. 
GNI04 show the effect that varying the value of $\alpha$ has on the
temperature structure of the gas, but finally adopt a model with 
$\alpha$~=~0.01 for their line emission predictions (Glassgold et al. 2007; 
Meijerink et al. 2008). A value of $\alpha =0.01$ is also chosen by 
d'Alessio et al. (1998, 1999, 2001) for their disk structure calculations 
(which includes the model used in this paper). This choice is is based on the 
connection between $\alpha$ and the disk accretion rate: 
$\dot{M_D} \propto \alpha\Sigma_D$, where $\Sigma_D$ is the integrated surface
density of the disk. 
With $\alpha =0.01$, which is also the value implemented in our models,
 X-ray heating is dominant in the disk corona, while accretion heating 
dominates in the inner regions. 

We finally note that the dependence of the heating rate on the square of the sound
speed implies a {\em linear} dependence on the gas temperature, which poses potential 
problems for the hot disk corona. In the regions where the gas is heated to high temperatures
by X-ray irradiation the functional form of accretion heating, as formulated above, fails
completely, preventing an equilibrium temperature from being found. In fact, X-ray irradiation in
the upper atmosphere of the inner regions of the disk heats the gas to
temperatures above the local escape temperature of the gas (see section~\ref{s:disp}). 
The gas in these regions is to be considered unbound and part of a
photoevaporative flow, hence not subject to the standard viscous accretion
law. For unbound gas we remove the accretion heating contribution altogether by setting 
the local $\alpha$  coefficient to zero in our models.

\section{The ionisation and temperature structure} \label{s:ion}

In this section, we present the temperature and ionisation structure of
the gas in the inner disk, focusing our attention on the disk corona.
We limit our discussion to radial distances $R \la 40$~AU, beyond 
which, in our model, sets a rough outer limit to the region where 
the effects of X-ray irradiation on the temperature and ionisation 
structure are more evident. We note, however, that in our models a low levels of ionisation 
is maintained out to a radius of $\sim$190~AU.

\begin{figure}
\plotone{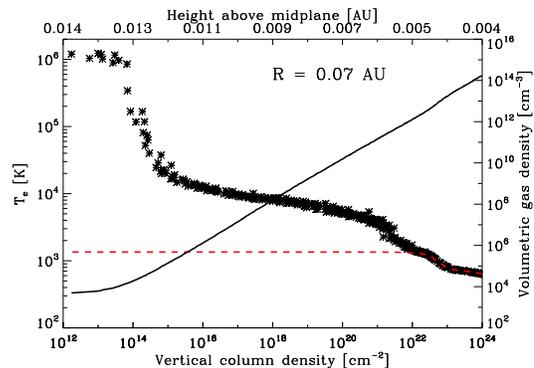}
\caption[]{Gas temperature structure (asterisks) as a 
function of gas column depth in the disk atmosphere at a radial distance of
0.07~AU. The dust temperature distribution of D'Alessio et al. (1999)
is plotted as the red dashed line. The black solid line shows the
volumetric gas density $[$cm$^{-3}]$ as a function of depth.}
\label{f:f1}
\end{figure}

Figure~\ref{f:f1} shows the temperature structure (asterisks) in the vertical
column calculated at $R$~=~0.07~AU, together with the the dust
temperature profile calculated by D'Alessio et al. (1999).  Also shown
is the volumetric gas density $[$cm$^{-3}]$ as a function of depth, the corresponding
geometrical height above the midplane is shown in the top x-axis.
The gaseous disk corona at these very small radii is fully ionised.
Since all elements are stripped of their outer electrons, cooling by
collisionally excited lines is inefficient. The balance between
heating by photoionisation and cooling by line emission and
recombination processes corresponds to a gas equilibrium temperature
of $\sim$10$^6$K.
 
At 0.07~AU, the jump in gas temperature occurs at a vertical column
density of $\sim 5 \times 10^{14}$~cm$^{-2}$, where the gas density is
$\sim 3 \times 10^{4} cm^{-3}$ and the ionisation
parameter, defined as $\xi$~=~L/nR$^2$, is log($\xi$)~$\sim$1.78.  
This transition
would look even sharper in a hydrostatically balanced model, since a
rise in the temperature would induce a drop in density in order to
mantain pressure equilibrium. In this case, any equilibria found in
the 10$^4$-10$^6$K range may in fact be unstable (Hatchett,
1976). This has already been noticed in accretion disk models around
X-ray binaries (e.g. Raymond 1993).
Alexander, Clarke \& Pringle (2004, ACP04) have also presented a model 
to illustrate the effects of X-ray heating on the structure 
of circumstellar disks. 
In that work, the density profile was truncated at the 
point where the density fell to $10^4$cm$^{-3}$, which is below the fully 
ionised region shown in our Figure~\ref{f:f1}. This should be kept in mind 
when comparing our temperature structure to theirs (as shown in their 
Figure~3), where the lack of the $\sim$10$^6$K temperature (fully ionised) 
region is not surprising and is not discrepant with our results. 

\begin{figure}
\plotone{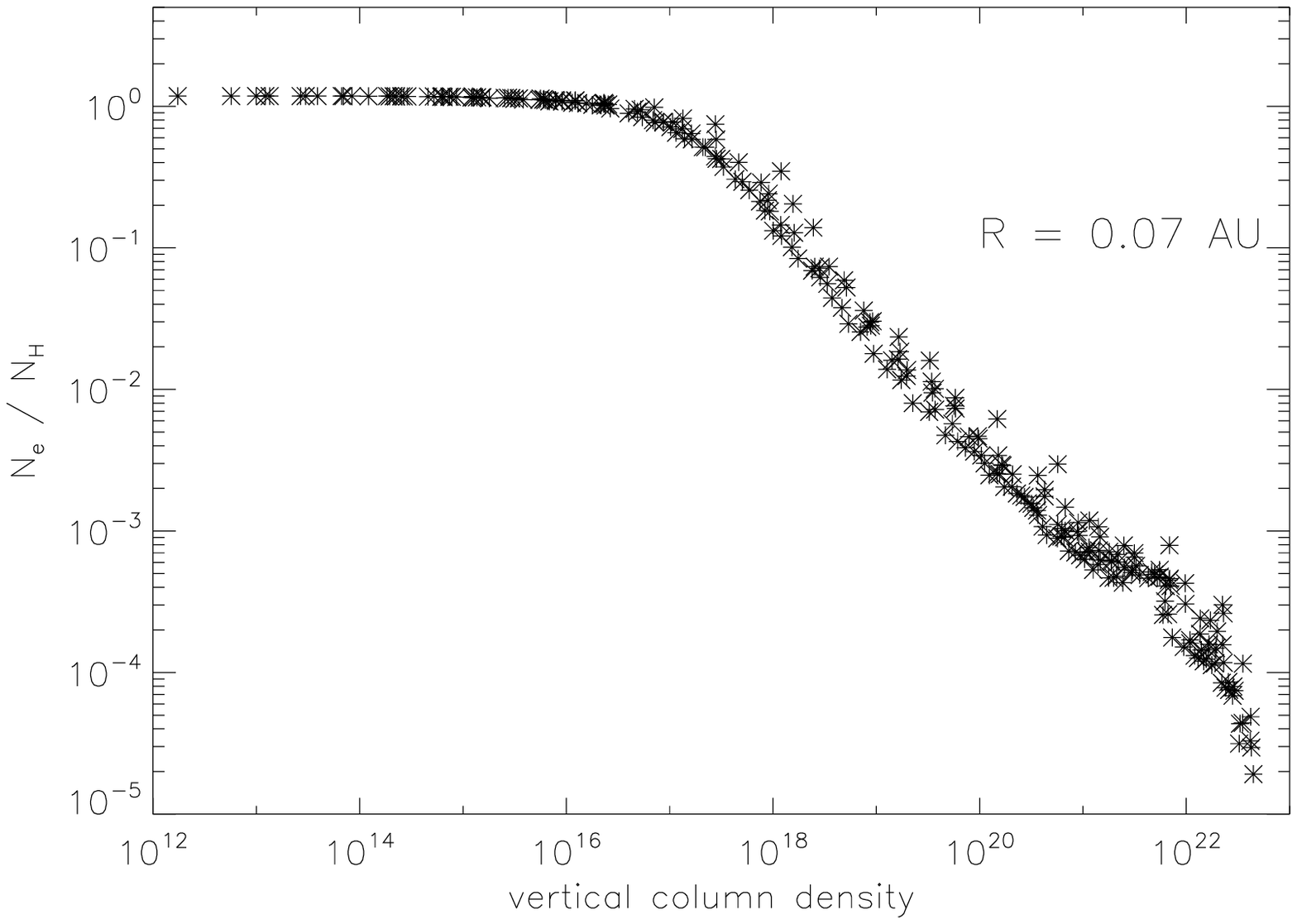}
\caption[]{Fractional 
abundance of electrons with respect to hydrogen nuclei plotted as a function 
of depth in the disk atmosphere at a radial distance of 0.07~AU.}
\label{f:f2}
\end{figure}

The electron density distribution, shown in Figure~\ref{f:f2},
indicates that the a high level of ionisation at 0.07~AU is maintained for
a column density of 10$^{17}$~cm$^{-2}$, after which it decays to
lower values.  A study of the thermal balance at this radius indicates
X-ray heating dominates for a column depth of 
$\sim$10$^{22}$~cm$^{-2}$, while heating by viscous accretion dominates
further down in the disk where it is balanced by dust-gas collisional
cooling.  In our calculations, the contribution from accretion heating
is set to zero in regions where X-ray irradiation heats up the gas to
temperatures above the local escape temperature (see
section~\ref{s:model}).  
At $R=0.07$~AU this occurs at a
column depth of $\sim$10$^{14.5}$~cm$^{-2}$, at the onset of the temperature
'discontinuity'.  
 
\begin{figure}
\plotone{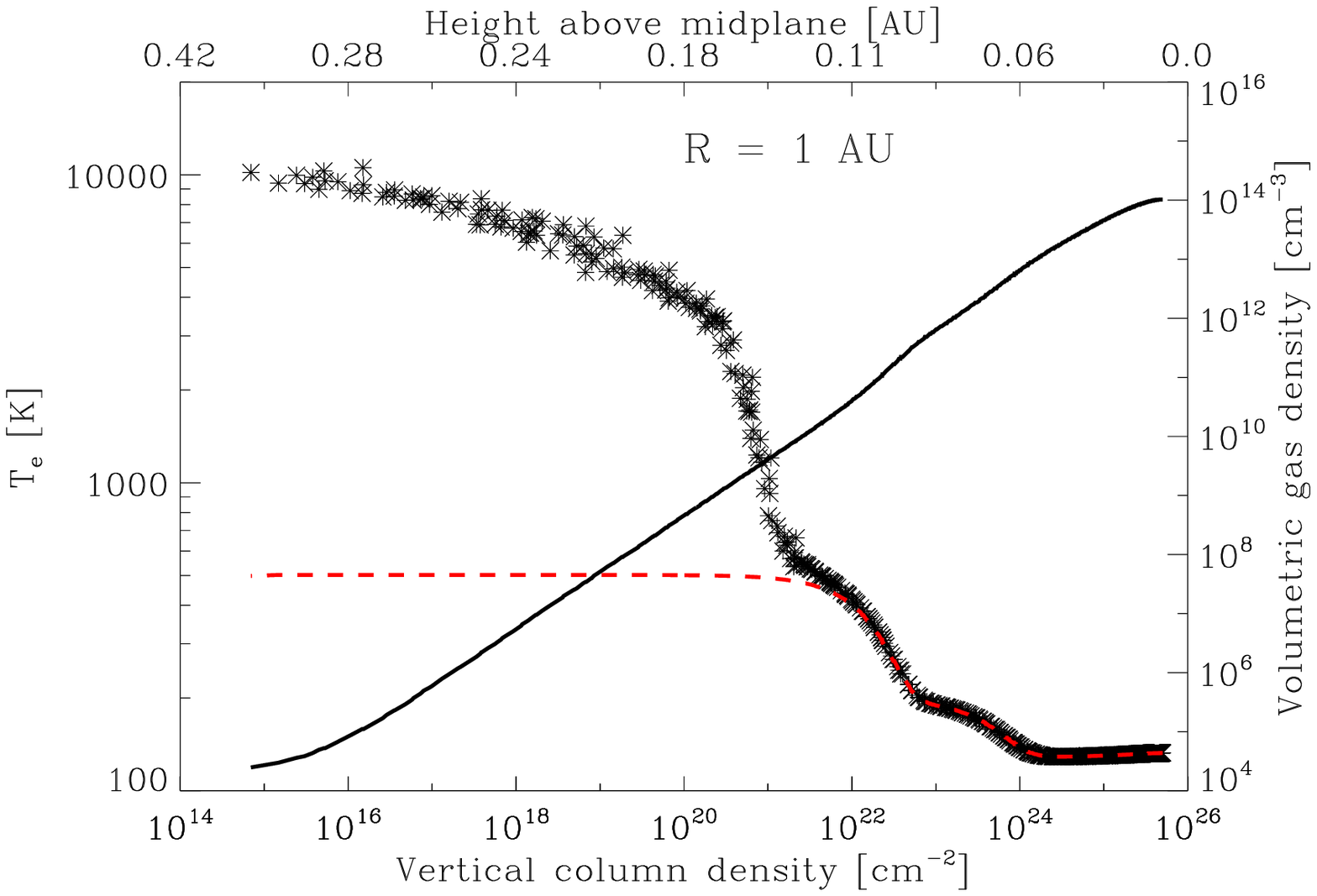}
\caption[]{Gas temperature structure (asterisks) as a function of depth 
in the disk atmosphere at a 
radial distance of 1~AU. The dust temperature distribution of
D'Alessio et al. (1999) is plotted as the red dashed line. The black
solid line shows the volumetric gas density $[$cm$^{-3}]$ as a
function of depth.}
\label{f:f4}
\end{figure}
\begin{figure}
\plotone{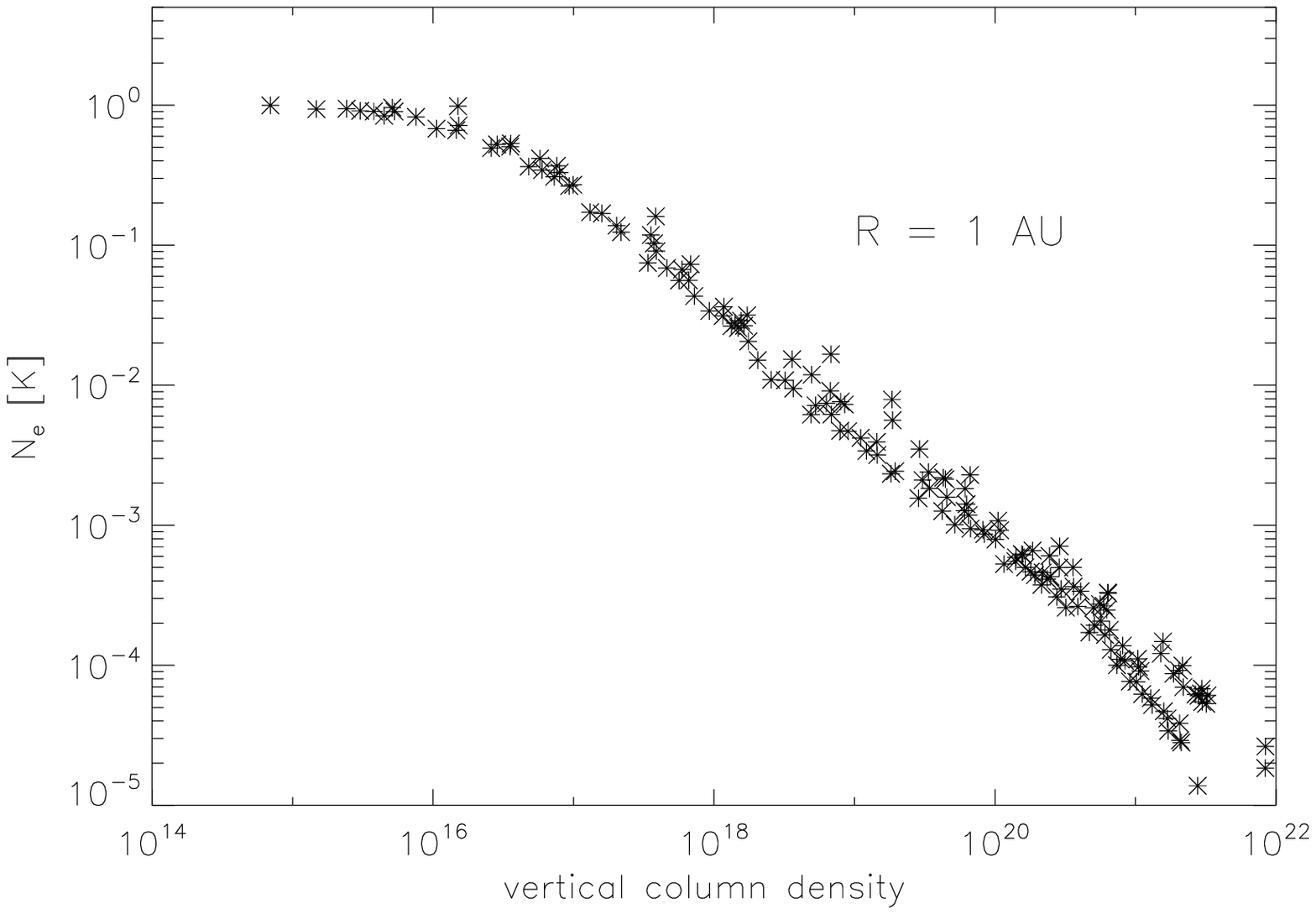}
\caption[]{Fractional abundance of electrons with respect to 
hydrogen nuclei plotted as a function 
of depth in the disk atmosphere at a radial distance of 1~AU.}
\label{f:f5}
\end{figure}

Figure~\ref{f:f4} shows the temperature and volumetric density
$[$cm$^{-3}]$ structures in the vertical column calculated at
$R$~=~1~AU, together with the D'Alessio et al.\ (1999) dust temperature
profile.  The coronal gas here appears to be only partially ionised
and the electron density, illustrated in Figure~\ref{f:f5}, reaches a
level of just over 10\% of the hydrogen density, but decreases very
rapidly with column depth.

\begin{figure}
\plotone{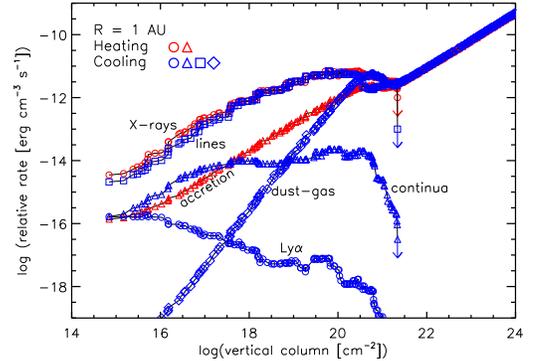}
\caption[]{Thermal balance as a function 
of depth in the disk atmosphere at a radial distance of 1~AU. 
The jaggedness of the lines is due 
to Monte Carlo statistical error. 
Please refer to the online version of this paper for a colour figure.
Red circles: heating by
X-ray photoionisation; red triangles: heating by viscous accretion; blue circles: 
cooling by Ly$\alpha$ emission; blue triangles: cooling by free-free and free-bound continuum 
emission from hydrogen and helium; blue squares: cooling by collisionally excited lines 
of metals; blue diamonds: cooling by dust-gas collisions.}
\label{f:f6}
\end{figure}

Our temperature structure at 1~AU is comparable to that found by
GNI04, allowing for differences in the model input parameters, and
bearing in mind that their calculation started at a column depth of
10$^{18}$~cm$^{-2}$. However, the cooling and heating contributions
calculated by the two models disagree somewhat and deserve further
explanation.  The various heating and cooling channels in our model at
1~AU are illustrated as a function of vertical column depth in
Figure~\ref{f:f6}, which can be compared with an analogous illustration
in Figure~3 of GNI04.  One conspicuous difference between the two
models is the accretion heating contribution, which is simply due to
the different choice of $\alpha$ (0.01 in this work and 1.0 in the 
particular model plotted in Figure~3 of GNI04).
Note that GNI04, also investigated models with lower values of $\alpha$
(see their Figure~2), and indeed used a model with $\alpha$~=~0.01, for their 
later work (Glassgold et al., 2007; Meijerink et al.; 2008), where emission 
line predictions are given. Unfortunately a plot showing the heating/cooling contributions from their models is only available for the $\alpha$~=~1 case.
Another important difference lies in the Ly$\alpha$ cooling contribution, 
which is only of secondary importance in our models but dominates the cooling
down to a vertical column of 10$^{21}$~cm$^{-2}$ in the GNI04
model. The dominant cooling mechanism in the upper disk atmosphere in
our model is metal line emission. The Ly$\alpha$ cooling in GNI04 was
overestimated because of an assumption of a thermal equilibrium population for
the n~=2 level. The ``Ly$\alpha$'' channel in the GNI04 model also
includes cooling by the fine structure lines of O~{\sc i}, while other
atomic coolants are ignored. In spite of the aforementioned
differences, the total cooling calculated by GNI04 is very similar to
that obtained in this work for the regions where the calculations
overlap and when the same $\alpha$ parameter is employed (Glassgold, 2008; 
private communication). 

As we move to increasingly larger radii, up to approximately
25-30~AU, our model disk structure remains qualitatively 
similar to that at 1~AU: high gas
temperatures are reached in the corona, followed by a steep temperature
decrease, with the gas and dust becoming thermally coupled at a vertical 
column depth of $\sim$5$\times$10$^{21}$cm$^{-2}$.  Beyond $\sim$35-40~AU,
due to geometrical dilution of the central radiation field, the effects of X-ray 
irradiation become less evident, however a low level of ionisation is maintained 
out to a radius of $\sim$190~AU. 
For T-Tauri stars born in clusters containing massive stars, like the Orion
Nebula Cluster, the dominant irradiation mechanism at large disk radii is 
probably provided by the interstellar or intracluster FUV and X-ray fields, 
which are not included in our current models. This is not true for stars born in 
smaller clusters like $\rho$ Ophiuchi and Taurus-Auriga, where indeed large disks 
($\sim$500-1000~AU) are found (Andrews \& Williams, 2007).

The maximum temperature reached by the disk corona at a given radial
distance is an immediate consequence of the local ionisation
parameter.  The fact that at 1~AU we do not see the $\sim$10$^6$K gas
layer is simply due to the fact that the density distribution of
d'Alessio et al. (1999) used in our models is truncated at a
volumetric density of $\sim$10$^4$. At 0.07~AU, the temperature jump
to $\sim$10$^6$K occurs at a log($\xi$)~=~1.78. The same value at 1~AU
implies a density of $\sim$150~cm$^{-3}$, which is below the cut-off.

The general results found in our models are similar to those found by 
Gorti \& Hollenbach (2008), in particular that the gas and dust are 
thermally de-coupled down to a column of $\sim10^{22} cm^{-2}$. 

\section{Predicted emission line spectrum}\label{s:lines}

As an aid to future observational campaigns, we have calculated the
emission line spectrum predicted by our model. Line luminosities are 
integrated out to a radius of 190~AU. In Tables~\ref{t:t1} and 
~\ref{t:t2} we list a subset of the strongest 
collisionally excited metal line and hydrogen
recombination lines, respectively, down to a luminosity of
$\sim$10$^{-8}$L$_{\odot}$, which corresponds to a flux of
$\sim$1.8$\times$10$^{-17}$erg/s/cm$^2$ at a distance of 140pc. 
The flux at a distance $D$, assuming a face-on orientation for the 
disk, is obtained by dividing the luminosities listed by the appropriate 
dilution factor (4$\pi D^2$). He~{\sc i} recombination lines are found to be 
rather weak, with the exception of He~{\sc i}~10.8$\mu$m which has a luminosity of 
8.8$\times$10$^{-7}$L$_{\odot}$, He~{\sc i}~7067{\AA} with luminosity of 3.2$\times$10$^{-8}$L$_{\odot}$
and  He~{\sc i}~3890{\AA}, He~{\sc i}~5877{\AA}, He~{\sc i}~6688{\AA} and 
He~{\sc i}~20.5$\mu$, all with luminosities of $\sim$2$\times$10$^{-7}$L$_{\odot}$.  
The full set of lines, including over 2000 transitions is available on
request from BE. The line list will be further improved and updated in
a future work, when the effects of simultaneous stellar and interstellar
FUV irradiation, as well as X-rays, will be considered. Some changes to
the emission line intensities are also expected when the X-irradiation 
will be studied in the context of a self-consistent
hydrostatic equilibrium calculation. 

Tables~\ref{t:t1} and~\ref{t:t2} provide an indication
of the lines that should be targeted for observations of the gaseous
phase of disk atmospheres.  A number of potential atomic and ionised
gas diagnostic lines should already be detectable with current
instrumentation. The $[$Ne~{\sc ii}$]$~12.8$\mu$m line, in particular,
has already been detected in T~Tauri disks with {\it Spitzer} by
Pascucci et al.\ (2007), Lahuis et al\. (2007) and Espaillat et
al. (2007), and with MICHELLE on Gemini North by Herczeg et
al. (2007).  The detections to date range from a few 10$^{-7}$ to
10$^{-5}$ L$_{\odot}$, in qualitative agreement with values
predicted by our model. The $[$Ne~{\sc iii}$]$~15.5$\mu$m line has so
far only been detected in Sz102 (Lahuis et al. 2007), implying a
$[$Ne~{\sc iii}$]$/$[$Ne~{\sc ii}$]$ of $\sim$0.06 for this object,
which is some 2.5 times larger than our model. As noted by 
Glassgold et al. (2007) the abundance of Ne$^{2+}$ is lowered by 
charge exchange reactions with neutral H, implying low ($\leq$0.1) 
ratios of $[$Ne~{\sc iii}$]$/$[$Ne~{\sc ii}$]$ in the X-ray heated gas.
Gorti \& Hollenbach (2008)
also discuss the effects of EUV irradiation which would enhance the 
predicted $[$Ne~{\sc ii}$]$ and $[$Ne~{\sc iii}$]$ line luminosities, 
and affect their ratio. They predict that the $[$Ne~{\sc iii}$]$/$[$Ne~{\sc ii}$]$
also depends on the details of the EUV spectrum, with hard EUV spectra naturally
producing the higher ratios. 

Table~\ref{t:t1}
clearly shows that a number of optical lines have fluxes that are 
significantly larger than the $[$Ne~{\sc ii}$]$~12.8$\mu$m fluxes. 
Unfortunately many optical 
and UV lines are not observable with current technology, as they are
 invisible against the stellar spectrum. In particular, we 
predict a luminosity of 3.3$\times$10$^{-5}$ for the Mg~{\sc i} line at 
4573$\AA$, for undepleted Mg abundances. This
corresponds to only a few \% of the continuum flux level for a
photosphere of 4000K and for an assumed resolving power of
50,000. While this line might be detectable under favourable
circumstances and where disk gas is not strongly depleted, we are
unaware of any such detections in the literature. 

The predicted absolute flux of the various lines depends, of
course, on the details of the disk structure, the gas abundances and
the ionising spectrum. As an example, Meijerink et al. (2008, MGN08),
show, for a fixed disk structure, the dependence of the predicted flux
of a number of fine structure and forbidden lines on the assumed X-ray
luminosity.  Furthermore, lines with very low critical densities like
the C~{\sc i} and C~{\sc ii} fine structure lines (see
Table~\ref{t:t4}) are also very sensitive to conditions in the outer
regions of the disk, where densities are small. These regions are
preferentially illuminated by the interstellar (or intracluster)
radiation fields which are not included in our current simulations.

Unfortunately, Table~\ref{t:t1} is not yet complete, as
reliable collision strengths were not available for all species, and a
number of less abundant elements were omitted from our calculations in
order to limit computational costs. In particular, the following
species were not included, due to lack of atomic data: Ne~{\sc i},
Mg~{\sc iii}, Si~{\sc i}, S~{\sc i}, Fe~{\sc i}, Fe~{\sc iv}, Fe~{\sc
ix}, Fe~{\sc xi}, Fe~{\sc xiii}. Due to its complexity, the Fe~{\sc
ii} ion was also omitted from the present calculations but will be
included in future models.

A further observational complication lies in the difficulty in
separating the emission from the disk, the wind, the accretion funnels
and the photosphere.  High resolution spectroscopy offers one
possibility, in which line profiles that are shifted and/or broadened
to different extents in the different emission regimes can be resolved
(e.g. Herczeg et al., 2007).  Following the suggestion in MGN08, in
future work we will present predicted profiles for a number of
potentially observable lines emitted from different regions of the
disk. An alternative way to distinguish lines formed in the upper
layers of the disk and those formed in the accretion funnel would
involve a comparison of transitions with different critical densities
(e.g. O~{\sc i}\,6302 with N$_{crit}$~=~1.5$\times$10$^{6}cm^{-3}$,
compared to O~{\sc i}\,63$\mu$m with N$_{crit}$~=~2$\times$10$^{4}cm^{-3}$).

We note that some of our predicted line intensities differ
somewhat from those calculated by MGN08. In most cases, the
differences may be explained by the fact that our work used a
different disk model for the underlying gas density distribution, and
different gas abundances. By way of comparison, we have also computed
a model that uses the same input parameters and disk model 
as those of MGN08 and we compare the 
predicted line fluxes to those of MGN08 in Table~\ref{t:t4}. Most
of the lines listed are in reasonable agreement, particularly when one
considers the two different approaches to the problem and the use of
different atomic data. MGN08 predict larger $[$Ne~{\sc ii}$]$~12.8$\mu$m 
line luminosities than in our model; this is explained by the fact that 
MGN08 do not include cooling by $[$Ne~{\sc ii}$]$ line, which, as 
 already noted by Gorti \& Hollenbach (2008) is found to be important. 
As a consequence the gas temperature in the $[$Ne~{\sc ii}$]$ emitting region
will be higher in MGN08's calculations, producing a larger
 $[$Ne~{\sc ii}$]$~12.8$\mu$m luminosity. 
The C~{\sc ii} fine-structure line at 158$\mu$m shows a larger discrepancy 
and deserves further
attention. Because the critical density of this line is low
(see Table~\ref{t:t4}), it will be collisionally quenched at even
modest depths into the disk, and will be preferentially emitted in the
less dense upper layers of the disk corona. MGN08 calculations,
however, were focused on the characterisation of lower layers in the
disk and stopped short of the corona, with the result that some of the
flux from the C~{\sc ii} bright regions is probably missed. 

A detailed comparison of the line luminosities predicted by Gorti \& Hollenbach 
(2008) to those predicted by our models is hindered by the  different parameters and approaches employed by 
the two models.
However we note that the  
$[$Ne~{\sc ii}$]$~12.8$\mu$m luminosity predicted by our models is a factor 
or 3.3 lower than that derived by their fiducial model (Model A, their Table~4). 
This can be explained by the fact that our density distribution is determined 
by the dust temperature structure. Indeed Gorti \& Hollenbach (2008) comment that 
when they assume a dust-determined temperature structure they obtain a factor of $\sim$4 
lower $[$Ne~{\sc ii}$]$ line luminosities. We also note that our predicted 
C~{\sc i} and C~{\sc ii} fine-structure lines are higher by a factor of $\sim$2 
than those given 
by model~A in Gorti \& Hollenbach (2008). As discussed above, these lines have 
low critical densities and are therefore extremely sensitive to the details of 
the density distribution in the emitting region, furthermore, we use a factor of $\sim$3 
higher elemental abundance for C. These two factors are probably responsible for the 
small discrepancy at hand.
Gorti \& Hollenbach (2008) use highly depleted values for the Si elemental abundance; 
this is the likely cause of the lower (factor of $\sim$10) Si~{\sc ii}~35$\mu$m line luminosities predicted 
by their models. Finally the factor of $\sim$3 lower O abundance used in their 
model is probably one of the causes of the small (factor of $\sim$1.5) discrepancy 
with the O~{\sc i}~63$\mu$m line luminosity.

\section{Disk dispersal by photoevaporation}
\label{s:disp}

Circumstellar disks are observed around the majority of young stars at
$\sim$10$^6$~yr (e.g. Lada \& Lada 2003). By ages of $\sim$10$^7$~yr,
however, only low-mass debris disks are generally observed
(e.g. Manning \& Sargent, 1997; Wyatt, Dent \& Greaves, 2003),
implying disk lifetimes of up to a few $\sim$10$^6$ yr. Indeed, recent
observations indicate that about 90\% of low mass stars lose their
primordial disks at 5-7 Myr (e.g. Haisch et al., 2001; Hartmann, 2005;
Hernand\'ez et al., 2007a). However, very few observations exist of
objects in transition between classical T~Tauri stars, which still
have disks of a few percent of a solar mass, and weak-lined T~Tauri
stars, which have dispersed nearly all of their circumstellar disks.
For example, Hernandez et al.~(2007b) reported only a 10\% frequency
of transitional disk candidates among the disk-bearing low mass stars
in their {\it Spitzer} observations of the Orion OB1 association.
This suggests that the mechanism responsible for the final stages of disk
dispersal must operate on relatively short timescales---perhaps as short as 
$\sim$10$^5$~yr (e.g. Duvert et al., 2000, Andrews \& Williams, 2005). 

Photoevaporative heating from both the central star and external
sources has been posited as a viable mechanism able to satisfy the
two-timescale nature of the problem.  A dramatic illustration of the
latter is the apparent photoevaporation of Orion ``proplyds'' in the
vicinity of OB stars (e.g. Johnstone et al. 1998, Adams et al., 2004; 
Fatuzzo \& Adams, 2008).  Evidence is also
accumulating that disk loss is more rapid in more O-star rich
clusters, as judged from observed disk frequencies in low-mass stars
(e.g. Albacete-Colombo et al. 2007, Guarcello et al. 2007, Mayne et
al. 2007).  We will return to photoevaporation by external sources of
radiation in future work, and concentrate here on the effect on the
disk of irradiation by the central T~Tauri star.

Photoevaporation by ultraviolet (UV) radiation from the central star
has been investigated by a number of authors (see Alexander, 2007; 
 Dullemond et al, 2007; Hollenbach, Yorke \&
Johnstone, 2000, for recent reviews). 
The original models did not couple photoevaporation
with viscous evolution, resulting in disk dispersal timescales
exceeding the observational estimates. Subsequently, the 'UV switch'
model of Clarke et al. (2001), which coupled photoevaporative mass 
loss to viscous evolution, was able to reproduce the two-timescale
behaviour implied by the observations. However a number of problems
still remain, as pointed out in Clarke et al. (2001).  In particular,
current models require that the Lyman continuum emission from the central star 
does not derive from accretion, an issue that is currently unclear (ACP04).
X-ray ionisation and heating of the disk corona provide an attractive
alternative engine to drive photoevaporative flows. The fact that  
weak-line T~Tauri stars, for which accretion has nearly ceased, 
 are known to be brighter in the X-ray than classical T-Tauri stars, 
which are still accreting, suggests that the production of X-ray is 
not linked to accretion. 
ACP04 estimated an
upper limit to the mass-loss driven by X-rays and found it to be at
best only comparable to that derived for ultraviolet evaporation.
They concluded, therefore, that X-ray heating is unlikely to be the
dominant disk dispersal mechanism.  ACP04 took into account the
effects of X-ray photoionisation and heating on the hydrostatic
structure of a typical T~Tauri circumstellar disk, but were 
limited by the fact that the radiative transfer problem could
not be solved self-consistently in 2D.

\begin{figure}
\plotone{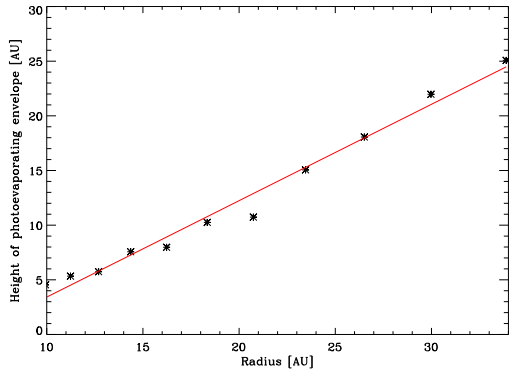}
\caption[]{Height of the photoevaporative envelope traced across the major photoevaporation region. 
The slope of line relating this quantity to the radial distance is 0.9. 
}
\label{f:f9}
\end{figure}

While our models do not yet iterate over the disk density structure,
which would be undoubtedly modified by X-ray irradiation, we do
calculate the irradiated thermal structure in 2D.  It is therefore
worthwhile to provide an estimate of the mass-loss rates and dispersal
timescales implied by our models. We estimate mass-loss rates by
locating, at each radial distance, the base of the photoevaporative
flow, defined as the height where the local gas temperature exceeds
the escape temperature of the gas, defined as $T_{es} = G m_H M_* / k R$, 
where $M_*$ is the stellar mass, $R$ is the radial distance. 
This height is illustrated as a function of radial distance in
Figure~\ref{f:f9}. The figure shows that, between 8 and 40~AU, the
height of the 
photoevaporating envelope follows a nearly linear increase with radial
distance from the star (slope $\sim$0.9).
To a first approximation, the mass
loss rate per unit area can be estimated as $\dot\Sigma_x = \rho c_s$, where
$\rho$ and $c_s$ are, respectively, the gas density and the sound
speed evaluated at the base of the flow.

\begin{figure}
\plottwo{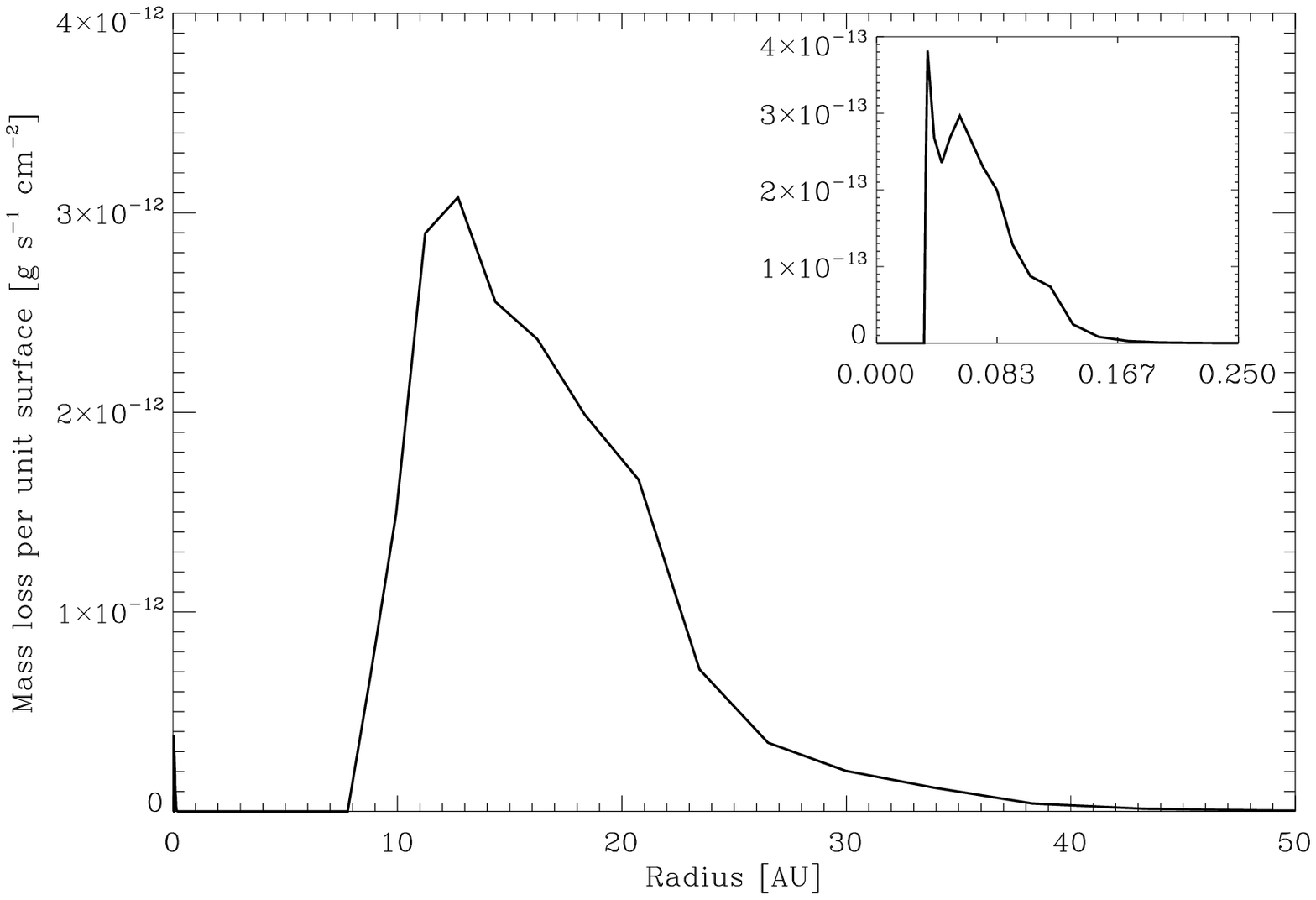}{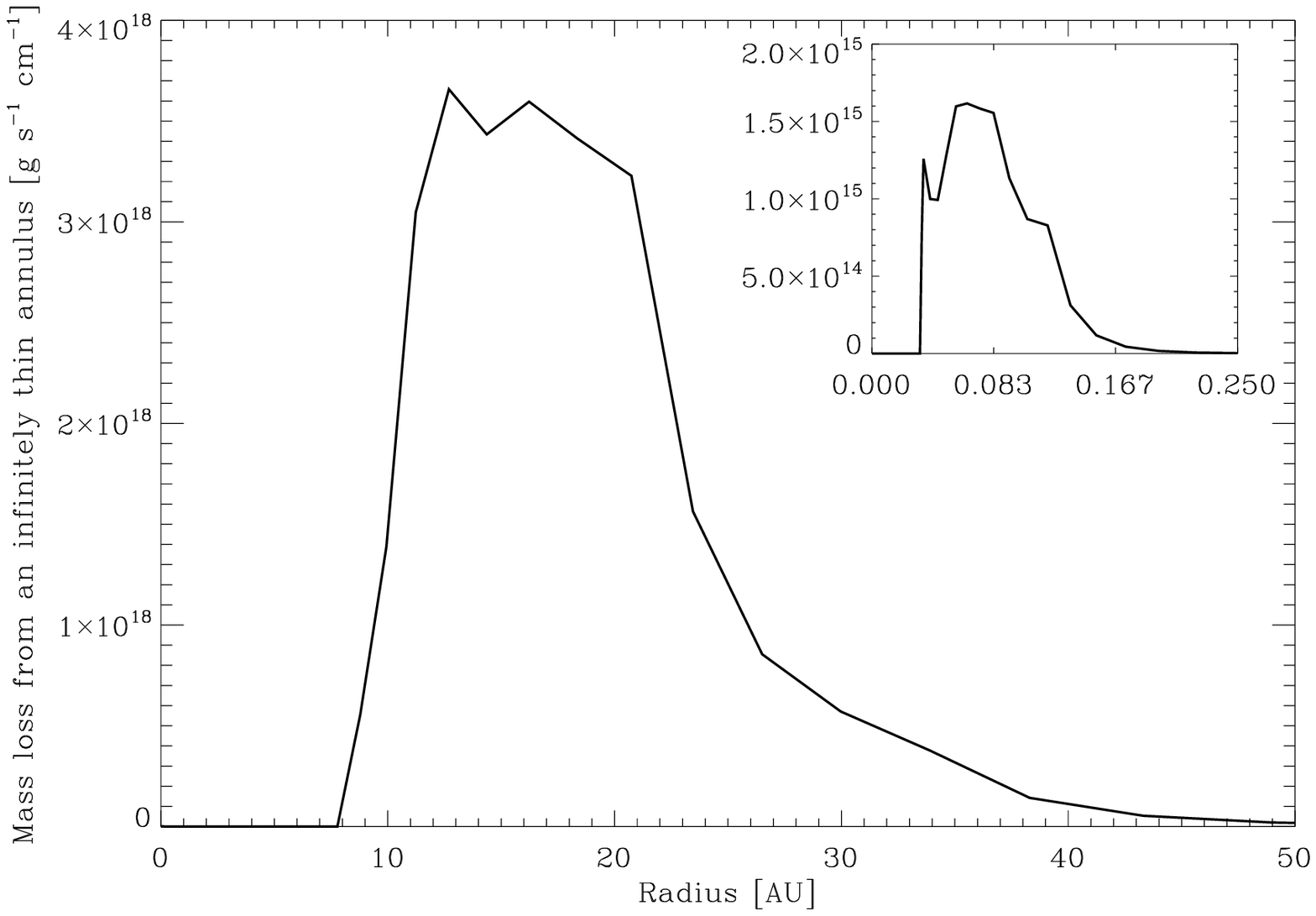}
\caption[]{Left: Mass loss per unit surface area 
as a function of the radial distance from the central star. 
Right: Mass loss from an infinitely thin annulus as a function of the radial distance from the central star. 
The insets show the 0.to 0.3~AU region enlarged for clarity. }
\label{f:f8}
\end{figure}

Figure~\ref{f:f8} (left panel) shows 
{$\dot\Sigma_X$}, the mass loss rate per unit area 
and (right panel) the mass loss rate from an infinitely thin annulus
 as a function of radial distance in the
disk. {$\dot\Sigma_X$} peaks at approximately 12~AU, near the gravitational 
radius, $r_g$ for gas with sound speed of $\sim 7 km/s$ ($T \sim 6000~K$). We use the 
definition of $r_g$ given by Hollenbach et al. (1994), where $r_g = G M_* / c_s$. 
The value at the peak is 3.1$\times$10$^{-12}~g\,s^{-1}\,cm^{-2}$, this is about a factor of
ten higher than the estimate of ACP04
(2.6$\times$10$^{-13}~g\,s^{-1}\,cm^{-2}$) and the value obtained by
Hollenbach et al. (1994) for UV photoevaporation
(3.9$\times$10$^{-13}~g\,s^{-1}\,cm^{-2}$). We note that both 
studies considered the case of a 1~M$_{\odot}$ star; however, since photoevaporation
rates scale only weakly with mass ($\sim M^{0.5}$ for EUV photoevaporation 
and almost linearly for X-ray photoevaporation), this is not sufficient to 
explain the order of magnitude difference with our rates. 
As discussed in the next 
section, our model has a number of limitations which may result in 
uncertainties in the derived photoevaporation rates. Taking into account
statistical uncertainties in the calculated temperatures alone, we find 
that our models are also consistent with peak values as low as 
$\sim$8$\times$10$^{-13} g\,s^{-1}\,cm^{-2}$, only a factor of $\sim$ 4 higher 
than previous estimates when corrected for the different stellar masses.

The effects of X-ray
photoevaporation are only significant between 8 and 40~AU. There is
also a small inner region between 0.03 and 0.16~AU, where, in spite of
the large escape temperatures, the intense X-ray field causes the disk
to lose mass at a rate of 3-4$\times$10$^{-13}~g\,s^{-1}\,cm^{-2}$. Given 
the small area at small radii, this does not contribute significantly to the 
mass loss.
The reason for the mass loss being mainly concentrated between 8-40~AU 
is that this region lies far enough from the central star for the 
escape temperatures to become sufficiently low and not so far that 
the effects of X-ray irradiation become unimportant. In other words, 
the steep drop in $\dot\Sigma_X$ inside 8~AU corresponds to the location 
where the sound speed at $\sim$10$^4$~K drops below the Keplerian speed. 
Any mass loss at smaller radii comes from coronal gas. In fact, 
for radial distances smaller than 8~AU, there is
also a thin layer of gas in the disk corona where X-ray irradiation
causes temperatures to rise above the local escape temperatures.
However, the contribution of this region to the total mass loss budget
is negligible. We also note here that by analogy with the X-ray heated 
accretion disk models of Begelman, McKee and Shields (1983) the wind region
might extend inward to about one tenth of the radius where the sound speed 
equals the escape speed. In that case the wind in Figure~\ref{f:f8} would begin 
at about 0.8~AU and the total mass loss rate would be increased. More recent 
works by Liffman (2003) and Font et al. (2004) also find that an EUV-driven 
wind is launched from well inside the gravitational radius (as defined by 
Hollenbach et al., 1994).

  The dispersal timescale
implied by our model is of a few Myr, for a 0.027~M$_{\odot}$ disk, 
and the total mass loss rate is of the order of 
\.M$_X \sim 10^{-8}$~M$_{\odot}$\,yr$^{-1}$. The disk
however, as well as being photoevaporated, is also accreting.  
Therefore, we can assume that the effects of X-ray photoevaporation
would not be felt until its viscous accretion rate, \.M$_{acc}$, has
fallen below \.M$_X$, at which point the disk would empty rapidly
within 10-20~AU.  Understanding the later evolution of the disk, after
the region within 10-20~AU has been emptied, requires us to determine
whether the disk beyond $\sim$20~AU ({\sc i}) has to accrete in to the
10-20~AU region and then be photoevaporated or ({\sc ii}) is more or
less fixed in density profile and then overtaken by the outward
propagating ring of mass loss. Behaviour of type {\sc i} (or {\sc ii})
results if the mass loss rate increases (decreases) as the hole size
increases. This question will be answered in a following study where
we plan to solve for the density structure and the 2D X-ray transport
self-consistently.

Our current calculations give
\.M$_X \sim 10^{-8}$~M$_{\odot}$\,yr$^{-1}$, implying that
disks with \.M$_{acc} \leq 10^{-8}$~M$_{\odot}$\,yr$^{-1}$ should disperse rapidly
and therefore be seldom observed. However this is in disagreement with
observations of a considerable number of systems in
the 0.5-1~M$_{\odot}$ range that are reported to have accretion rates lower
than $10^{-8}$~M$_{\odot}$\,yr$^{-1}$. Indeed accretion rates as low
as $10^{-9}$~M$_{\odot}$\,yr$^{-1}$ are seen in the $-0.2 < \log M_* <
0$ mass bin in Figure~8 of Gregory et al. (2006), which, collecting
data from a number of authors, shows the 
correlation between observed mass accretion rates and stellar
mass. This indicates that our predicted \.M$_X \sim
10^{-8}$~M$_{\odot}$\,yr$^{-1}$ is probably too high, and may point to
a weakness of our model (see discussion in the next Section). 

\subsection{Limitations of our models}
\label{s:lims}

A number of important approximations were used 
in our models, which may lead to uncertainties in the estimates of the 
photoevaporation rates.

One major caveat of our model, already mentioned in this paper, is
that the influence of X-ray photoionisation and heating on the disk
density structure has been ignored. The high gas temperatures reached
in the inner regions of the disk would cause an expansion of the gas
in the z-direction, as shown, for example, by ACP04. Indeed, their
lower value of mass loss rate is perhaps caused by a larger
attenuation of the X-ray field due to the 'puffed up' inner disk
material that is not accounted for by our models. A hot and inflated
inner disk could in fact cause a shadowing effect of the regions at
larger radial distances, reducing the effects of X-ray
photoevaporation there. The magnitude of this effect can only be
properly estimated via the simultaneous solution of the 2D radiative
transfer problem and the 1D hydrostatic equilibrium in the
z-direction, which is the aim of a follow-up paper. 

Further uncertainties are introduced by the use of the ``$\rho \times c_s$'' 
approximation in the calculation of photoevaporation rates. In 
Section~\ref{s:disp} we discuss how the base of the photoevaporative 
envelope is defined as the height where the local gas temperature exceeds
the escape temperature of the gas. While this is generally a very good 
approximation in cases where the temperature and density vary sharply 
at the ionisation front, as for EUV photoevaporation (e.g. 
Hollenbach et al., 1994), in the case of X-ray photoevaporation presented here, 
the distinction between the bound and unbound gas is less clear. In other
words, a small error in the determination of temperatures in our models, 
would lead to much larger uncertainties in the density, and hence in the 
estimated photoevaporation rates. 

We conclude that while we have shown that X-rays have the potential to drive 
an outflow, our derived rates should be treated with caution and 
we defer firm conclusions to future work.


\section{Summary}\label{s:summary}

We have presented a 2D photoionisation and dust radiative transfer
calculation of a typical T~Tauri protoplanetary disk irradiated by
1.5$\times$10$^{7}$~K (kT~=~1.36 keV) X-ray radiation field from the central
pre-main sequence star.

A detailed characterisation of the temperature structure of the upper
disk corona was provided, which uncovered the existence of a layer of
highly ionised, hot gas in the inner disk regions. The maximum
temperatures reached by the gas in the upper corona were mapped across
the radial dimension of the disk, showing peaks exceeding 10$^6$K in
the very inner regions. The hotter component of the gas in the 
inner disk and at radii
between $\sim$8 to $\sim$40~AU was shown to be unbound, and probably
part of a photoevaporative flow.

To aid future efforts to observe the hot/warm gas component of T~Tauri
disks, we provide a substantial list of predicted emission line
luminosities, many of which should already be detectable with current
instrumentation.

The temperature structure of the irradiated disk implies a total mass
loss rate of $\sim10^{-8}$~M$_{\odot}$\,yr$^{-1}$ and a
disk dispersal timescale of a few Myr for a disk mass of
0.027~M$_{\odot}$, which is 
the mass of the disk structure model we employed. We discuss the limitations of our model and
highlight the need for further calculations that solve the 2D
radiative transfer problem simultaneously with the 1D hydrostatic
equilibrium of disk structure in the polar direction.

\section*{Acknowledgments}

We extend our warmest thanks to Al Glassgold, for insightful
discussion and a thorough assessment of our results. We also thank the 
anonymous referee and the editor Eric Feigelson for helpful comments that added to the clarity of the paper and the interpretation of the results. We thank Paola
D'Alessio for providing us with the electronic data for the gas
density distribution in the disk. We also thank Bruce Draine for help
with the X-ray dust opacity calculations. We finally extend our thanks
to Roger Wesson for helpful discussion. JJD was supported by the
Chandra X-ray Center NASA contract NAS8-39073 during the course of
this research.  BE was partially supported by {c\it Chandra} grants
GO6-7008X and GO6-7009X.The simulations were partially run on the Cosmos 
(SGI altix 4700) supercomputer at DAMTP in Cambridge. Cosmos is a 
UK-CCC facility which is supported by HEFCE and STFC.

\begin{table*}
\caption[]{Collisionally Excited Lines
in order of decreasing line luminosity. 
The flux at a distance $D$, assuming a face on orientation of the disk,
 is obtained by dividing the line luminosities by 4$\pi\,D^2$.}
\label{t:t1}
\begin{center}
\begin{tabular}{lcc|lcc}
\hline
Species & Wavelength  & Luminosity & Species & Wavelength  & Luminosity \\
        &  $[$A$]$    & $[$L$_{\odot}]$ &        &  $[$A$]$    & $[$L$_{\odot}]$\\ 
\hline
MgI   &     4573.18    &    3.3e-05	&  SIII   &    335008.    &   6.1e-08	   \\
SiII  &      348189.   &    1.4e-05     &  MgIV   &    44883.3    &   5.8e-08	   \\
OI    &    631712.     &    6.1e-06     &  OI     &  5578.89      &   5.5e-08	   \\
OI    &    6302.03     &    3.5e-06     &  SiII   &    2335.32    &   5.0e-08	   \\
MgII  &      2796.35   &    2.6e-06     &  NI     &  10400.6      &   4.2e-08	   \\
MgI   &     2852.14    &    1.6e-06     &  OIII   &    518135.    &   4.1e-08	   \\
MgII  &      2803.53   &    1.3e-06     &  CII    &   2328.84     &   3.9e-08	   \\
OI    &    6365.53     &    1.1e-06     &  SiII   &    2344.92    &   3.5e-08	   \\
SII   &     4069.76    &    1.0e-06     &  CII    &   2325.40     &   3.3e-08	   \\
CII   & 1.57729e+06    &    9.6e-07     &  OIII   &    5008.24    &   3.2e-08	   \\
CI    &    9852.99     &    9.5e-07     &  CI     &  2968.08      &   3.0e-08	   \\
SII   &     6732.69    &    8.7e-07     &  SIII   &    9070.05    &   2.9e-08	   \\
NeII  &      128156.   &    8.7e-07     &  NII    &   6549.86     &   2.7e-08	   \\
CI    &    8729.51     &    4.6e-07     &  CII    &   2324.21     &   2.4e-08	   \\
SII   &     6718.31    &    4.1e-07     &  NI     &  10410.5      &   2.1e-08	   \\
CI    &3.70370e+06     &    3.9e-07     &  FeXIII &      10746.2  &   1.9e-08   \\
CI    &    9826.85     &    3.2e-07     &  OII    &   3727.09     &   1.9e-08	   \\
OI    &1.45560e+06     &    3.0e-07     &  OIII   &    883392.    &   1.8e-08	   \\
SII   &     10323.3    &    2.9e-07     &  NI     &  10410.0      &   1.8e-08	   \\
SII   &     4077.51    &    2.7e-07     &  FeVI   &    147710.    &   1.6e-08	   \\
SII   &     10289.5    &    2.6e-07     &  FeVI   &    195580.    &   1.6e-08	   \\
CII   &     2326.11    &    1.9e-07     &  NI     &  3467.49      &   1.4e-08	   \\
SII   &     10339.2    &    1.8e-07     &  NIII   &    573394.    &   1.3e-08	   \\
SIII  &      9532.25   &    1.6e-07     &  NI     &  10401.1      &   1.3e-08	   \\
NeIII &       155545.  &    1.5e-07     &  CI     &  2965.70      &   1.3e-08	   \\
SIV   &     105108.    &    1.4e-07     &  SiII   &    1816.93    &   1.2e-08	   \\
CI    &6.09756e+06     &    1.3e-07     &  NeIII  &     3869.85   &   1.2e-08	   \\
MgI   &     4564.67    &    9.6e-08     &  FeXI   &    60823.6    &   1.2e-08	   \\
SIII  &      187056.   &    8.4e-08     &  FeXI   &    2649.46    &   1.1e-08	   \\
NII   &     6585.27    &    8.4e-08     &  SiII   &    2335.12    &   1.1e-08	   \\
SII   &     10373.3    &    7.3e-08     &  OIII   &    4960.29    &   1.1e-08	   \\
CII   &     2327.64    &    7.3e-08     &  FeVI   &    123107.    &   1.0e-08     \\
\hline                                     
\end{tabular}                              
\end{center}
\end{table*}     

\begin{table*}
\caption[]{Hydrogen Recombination Lines
The flux at a distance $D$, assuming a face on orientation of the disk, 
is obtained by dividing the line luminosities by 4$\pi\,D^2$.}
\label{t:t2}
\begin{center}
\begin{tabular}{ccc|ccc}
\hline
Transition & Wavelength  & Luminosity & Transition & Wavelength  & Luminosity \\
           &  $[$A$]$    & $[$L$_{\odot}]$ &        &  $[$A$]$    & $[$L$_{\odot}]$\\ 
\hline
 2-1   &     1215.68  &   3.0e-05  &       9-2  &     3836.51  & 1.2e-07 \\
 3-2   &     6564.70  &   2.6e-06  &       9-3  &     9231.60  & 4.0e-08 \\
 4-2   &     4862.74  &   9.5e-07  &       9-4  &     18179.2  & 1.7e-08 \\
 4-3   &     18756.3  &   3.2e-07  &      10-2  &     3799.01  & 1.0e-07 \\
 5-2   &     4341.73  &   4.7e-07  &      10-3  &     9017.44  & 3.3e-08 \\
 5-3   &     12821.7  &   1.6e-07  &      10-4  &     17366.9  & 1.4e-08 \\
 5-4   &     40522.8  &   7.8e-08  &      11-2   &    3771.74  & 9.6e-08 \\      
 6-2   &     4102.94  &   2.9e-07  &      11-3   &    8865.27  & 2.9e-08 \\      
 6-3   &     10941.2  &   9.8e-08  &      11-4   &    16811.2  & 1.2e-08 \\      
 6-4   &     26258.8  &   4.6e-08  &      12-2   &    3751.25  & 8.5e-08 \\      
 6-5   &     74598.8  &   2.5e-08  &      12-3   &    8752.93  & 2.6e-08 \\      
 7-2   &     3971.24  &   2.0e-07  &      12-4   &    16411.7  & 1.1e-08 \\      
 7-3   &     10052.2  &   6.7e-08  &      13-2   &    3735.47  & 7.7e-08 \\      
 7-4   &     21661.3  &   3.1e-08  &      13-3   &    8667.45  & 2.3e-08 \\      
 7-5   &     46537.9  &   1.6e-08  &      13-4   &    16113.8  & 1.0e-08 \\      
 7-6   &     123719.  &   9.8e-09  &      14-2   &    3723.03  & 6.8e-08 \\      
 8-2   &     3890.19  &   1.5e-07  &      14-3   &    8600.81  & 2.1e-08 \\      
  8-3  &     9548.65  &   5.0e-08  &      14-4   &    15884.9  & 9.1e-09 \\      
  8-4  &     19450.9  &   2.2e-08  &      15-2   &    3713.06  & 6.1e-08 \\      
  8-5  &     37405.7  &   1.2e-08  &      15-3   &    8547.78  & 1.8e-08 \\      
       &              &            &      15-4   &    15705.0  & 8.1e-09 \\      
\hline
\end{tabular}
\end{center}
\end{table*}

\begin{table}
\caption[]{Comparison of predicted line luminosities with MGN08}
\label{t:t4}
\begin{center}
\begin{tabular}{lcccc}
\hline
 Species & Wavelength  & N$_{crit}$  &\multicolumn{2}{c} {Luminosity $[$L$_{\odot}]$}   \\
         & $[$A$]$     & $[cm^{-3}]$ &   This Work    & MGN08 \\ 
\hline
OI    &    631712.     & 2.1e4  &    2.2e-05 &  3.8e-05\\
CI    &  6.09756e+06   & 1.7e0  &    8.5e-06 &  2.6e-07\\
CI    &  3.70370e+06   & 9.5e0  &    8.2e-06 &  9.2e-07\\
OI    &    6302.03     & 1.5e6  &    7.7e-06 &  7.5e-06\\
CII   &  1.57729e+06   & 4.9e1  &    3.0e-06 &  5.6e-09\\
CII   &    2326.11     & 1.4e9  &    1.1e-06 &  1.0e-06\\
NeII  &    128156.     & 5.9e5  &    9.5e-07 &  3.7e-06\\
CI    &    9826.85     & 1.4e4  &    8.9e-07 &  3.7e-06\\
OI    &    5578.89     & 9.5e7  &    4.9e-07 &  4.2e-07\\
NeIII &     155545.    & 2.0e5  &    1.9e-07 &  4.2e-07\\
\hline
\end{tabular}
\end{center}
\end{table}




\end{document}